\documentclass[aps]{revtex4-1}

\usepackage{amsfonts}
\usepackage{amsmath}
\usepackage{amsbsy}
\usepackage{graphicx}
\usepackage{subfigure}
\usepackage{xcolor}

\renewcommand{\vec}[1]{\mbox{\boldmath$\mathrm{#1}$}}

\def\ind#1{{_{\mathrm{#1}}}}

\begin{document}

\title{Finite-size effects on the magnetoelectric response of field-driven ferroelectric/ferromagnetic chains}

\author{Chenglong Jia$^1$, Alexander Sukhov$^{1,2}$, Paul P. Horley$^3$ and Jamal Berakdar$^1$}

\affiliation{$^1$Institut f\"ur Physik, Martin-Luther Universit\"at Halle-Wittenberg, 06120 Halle, Germany \\
$^2$Max Planck Institute of Microstructure Physics, Weinberg 2, 06120 Halle, Germany\\
$^3$Centro de Investigati\'{o}n en Materiales Avanzados, S.C. (CIMAV), Chihuahua/Monterrey, 31109 Chihuahua, Mexico}


\begin{abstract}
We study theoretically the coupled  multiferroic dynamics of one-dimensional ferroelectric/ferromagnet chains driven by  harmonic magnetic and electric fields \textcolor{black}{as a function of the chain length.  A linear magnetoelectric coupling is dominated by the spin-polarized screening charge at the interface.} We performed  Monte-Carlo simulations and calculations based on the coupled Landau-Lifshitz-Gilbert and Landau-Khalatnikov equations showing that the net magnetization and the total polarization of thin heterostructures, \textcolor{black}{i.e. with up to ten ferroelectric and ferromagnetic sites counted from the interface,} can be \textcolor{black}{completely} reversed by external electric and magnetic fields, respectively. \textcolor{black}{However, for larger system solely a certain magnetoelectrical control can be achieved.}
\end{abstract}

\maketitle

\section{Introduction}
\emph{Multiferroics}, i.e.\ materials in which ferroelectric (FE) and  ferromagnetic (FM) orders coexist, have attracted considerable attention in recent years \cite{multiferroics}. The strong magnetoelectric (ME) coupling   in these materials renders possible manipulation  by external magnetic and electric fields\cite{Lo04,Ki03,Bo05,We07,Ts06}. This  brings  about entirely new concepts in the design of next generation devices. Currently studied multiferroics are of  two kinds: single-phase systems and two-phase systems. Single phase multiferroics, such as RMnO$_3$ (\textcolor{black}{e.g.} R=Tb, Dy) \cite{Ki03,RMnO3} have successfully been exploited to control the magnetic order by electrical means and vice versa. A number of interesting phenomena  such as an electrically controllable persistent spin current \cite{JB-1} and a modified Datta-Das spin-field-effect transistor \cite{JB-2} associated with the two dimensional electron gas formed at the interface of multiferroic oxides was  reported. However, the small electric polarization and the low transition temperature  in single-phase multiferroics\cite{RMnO3} are adverse for  applications. On the other hand, two-phase multiferroics \cite{multiferroics,twophase}, including artificially synthesized ferroelectric and ferromagnetic materials, are very promising as a room-temperature ME system in which large ferroic orders coexist.

Very recently, the dynamic response to applied external fields in a multiferroic chain with a linear ME coupling induced by electrostatic screening at the ferroelectic/ferromagnet interface was theoretically investigated \cite{SJHB}.  It \textcolor{black}{was} demonstrated that for material parameters corresponding to BaTiO$_3$/Fe the \textcolor{black}{total} polarization and the \textcolor{black}{net} magnetization are controllable by external magnetic and electric fields, respectively. In this work we study the size-dependence of the multiferroic dynamics of the electric polarization and of the magnetization in several heterostructures. Calculations are performed within the framework of the coupled Landau-Khalatnikov \cite{LKh} and \textcolor{black}{zero Kelvin} Landau-Lifshitz-Gilbert equations \cite{LLG}. We find  that the polarization (magnetization) are magnetically (electrically) switchable for chains with a small size.

\begin{figure}[b]
\centering
\includegraphics[width=0.5\textwidth]{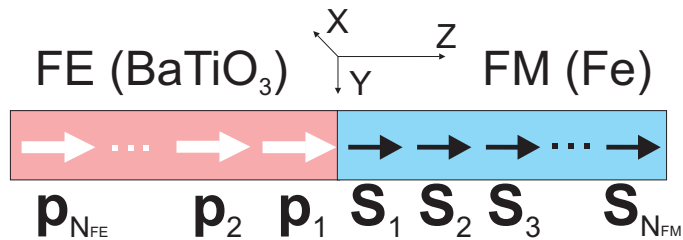}
\caption{Multiferroic chain consisting of $N_{FE}$ FE polarization sites and $N_{FM}$ localized FM moments. }
\label{fig1}
\end{figure}

\section{Theoretical formulation}
As sketched in Fig.\ref{fig1},  we consider a two-phase multiferroic chain consisting of
 an artificially grown ferroelectric-ferromagnetic heterostructure such as BaTiO$_3$/Fe.
  This  system is similar to that in which the magnetoelectric effect was recently predicted theoretically\cite{SJHB,Du06,Fe08} and realized experimentally\cite{Sa07}.  The total energy of the FE/FM-system in a general case falls into three parts \cite{SJHB}

\begin{eqnarray}
\displaystyle F_{\Sigma}=F_{\mathrm{FE}}\, + F_{\mathrm{FM}} + E_{\mathrm{c}},
\label{eq_1}
\end{eqnarray}
The ferroelectric energy contribution reads
\begin{eqnarray}
\label{eq_2}
F_{\mathrm{FE}}=a^3P\ind{S}&\sum_{j=1}^{N_{\mathrm{FE}}}\Big( \frac{\alpha_{\mathrm{FE}}P\ind{S}}{2}\,\vec{p}_j^2 + \frac{\beta_{\mathrm{FE}}P^3_{\mathrm{S}}}{4}\,\vec{p}_j^4+
   \frac{\kappa\ind{FE}P\ind{S}}{2}\,\big(\vec{p}_{j+1} - \vec{p}_{j}\big)^2 - \vec{p}_j\cdot\vec{E}(t)\Big),
\nonumber\\
\end{eqnarray}
and the ferromagnetic energy part is
\begin{eqnarray}
\label{eq_3}
   F_{\mathrm{FM}}=&\sum_{i=1}^{N_{\mathrm{FM}}}\Big( -J \,\vec{S}_i\cdot\vec{S}_{i+1}-D \,(S_i^{\mathrm{z}})^2
   -\vec{\mu}_i \,\vec{S}_i\cdot \vec{B}(t)\Big),
\end{eqnarray}
where $\vec{E}(t)$ and $\vec{B}(t)$ are the time-dependent applied external electric and magnetic fields, respectively. Various parameters and the values of reduced $\vec{p}_j$ and $\vec{S}_i$ entering eqs. (\ref{eq_2}) and (\ref{eq_3}) are discussed in details in Ref. \cite{SJHB}. In general there are several sources for the  ME coupling. Here the ME coupling arises due to the accumulation of spin-polarized electrons at the FE/FM interface when the FE subsystem is polarized\cite{Cai09}. Since the electric field does not penetrate into the bulk of FM metals and the induced electric charges decay exponentially away from the interface, the ME effect is  confined to a depth on the order of atomic dimensions from the surface.  Here, we consider an ideal FM metal, the exchange interaction between the induced surface magnetization and the FM part is thus limited  to the first site only. The linear ME-coupling energy is given as

\begin{eqnarray}
E_{\mathrm{c}}=\lambda \vec{p}_1\cdot \vec{S}_1.
\label{eq_4}
\end{eqnarray}
with $\lambda$ being the coupling strength.

The magnetization dynamics proceeds according to the \textcolor{black}{zero Kelvin} Landau-Lifshitz-Gilbert\cite{LLG} (LLG) equation of motion
\begin{eqnarray}
\label{eq_5}
   \displaystyle \frac{d\vec{S}_i}{dt} =& - \frac{\gamma}{1+\alpha^2\ind{FM}} \left[\vec{S}_i\times \vec{H}_i^{\mathrm{FM}}(t)\right]
   -\frac{\alpha\ind{FM} \gamma}{1+\alpha^2\ind{FM}} \left[\vec{S}_i\times\left[\vec{S}_i\times\vec{H}_i^{\mathrm{FM}}(t)\right]\right],
\end{eqnarray}
where $\gamma=1.76\cdot10^{11}~$ $(T\ s)^{-1}$ is a gyromagnetic ratio, $\alpha\ind{FM}$ a Gilbert damping parameter. The total effective field acting on $\vec{S}_i$ is defined as $\displaystyle{\vec{H}_i^{\mathrm{FM}}(t) = -\frac{1}{\mu_{i}}\frac{\delta F\ind{\Sigma}}{\delta \vec{S}_i}}$.\\ 
The polarization dynamics is governed by the Landau-Khalatnikov (LKh) equation as it was considered in the literature \cite{LKh} for ferroelectric  materials
\begin{eqnarray}
\displaystyle \gamma_{\nu}P\ind{S} \frac{d \vec{p}_j}{dt} = \vec{H}_j^{\mathrm{FE}}= -\frac{1}{P\ind{S}}\frac{\delta F\ind{\Sigma}}{\delta \vec{p}_j},
\label{eq_8}
\end{eqnarray}
where $\gamma_{\nu}$ is the viscosity constant.
Numerically both eqs. (\ref{eq_5}) and (\ref{eq_8}) are solved in reduced units, renormalizing the energy (\ref{eq_1}) over double anisotropy strength $2D$. Then, the dimensionless time in  both equations is $\tau=\omega\ind{A}\,t=\gamma\, B\ind{A}\,t=\gamma \, 2D\, t/\mu\ind{S}$ and the reduced effective fields are $\vec{h}_i^{\mathrm{FM}}(\tau)=\vec{H}_i^{\mathrm{FM}}(\tau)/B\ind{A}$, $\vec{h}_j^{\mathrm{FE}}=\vec{H}_j^{\mathrm{FE}}/(\gamma \gamma\ind{\nu}P\ind{S}B\ind{A})$.

\begin{figure}[t]
\begin{center}
\includegraphics[scale=0.5]{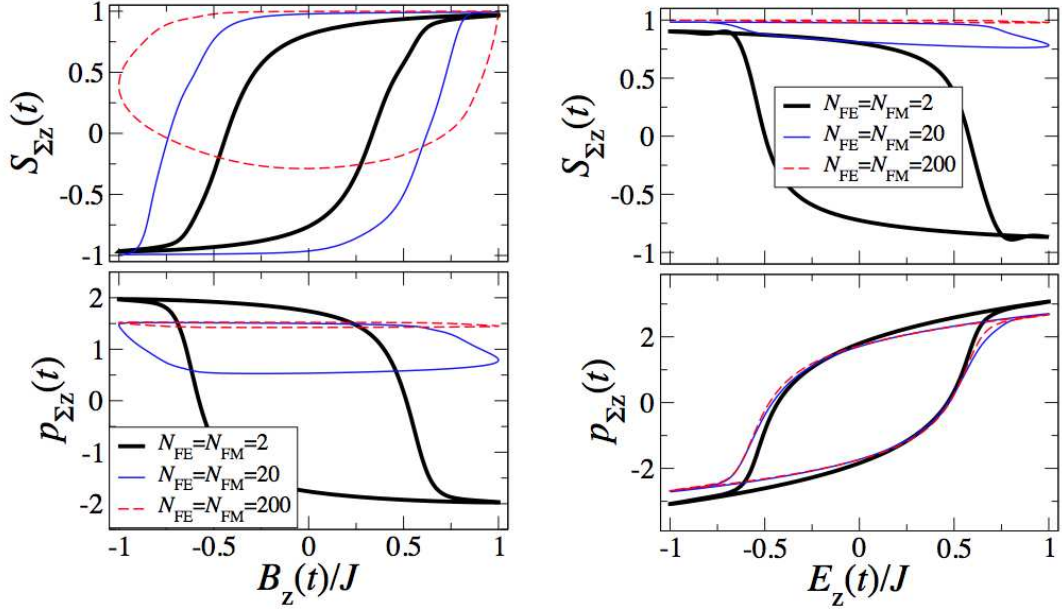}
\label{fig2}
\caption{The hysteresis loops of the reduced total polarization/magnetization of different ME chains in the presence of a harmonic external magnetic field $B_z(t) = B_{0z} \cos \omega t$ (\emph{left})  or electric field $E_z(t) = E_{0z} \cos \omega t$ (\emph{right}). The parameters are chosen such, that $\lambda = J$, $\omega = 3.61\times 10^{11}\  s^{-1}$, $\alpha_{FM} =0.5$. Amplitudes of external fields expressed in the energy units are $B_{0z} = J$ and $E_{0z} = 0$ (\emph{left}) and  $B_{0z} = 0$ and $E_{0z} = J$ (\emph{right}).}
\end{center}
\end{figure}
l
\begin{figure}[t]
\centering
\includegraphics[scale=0.5]{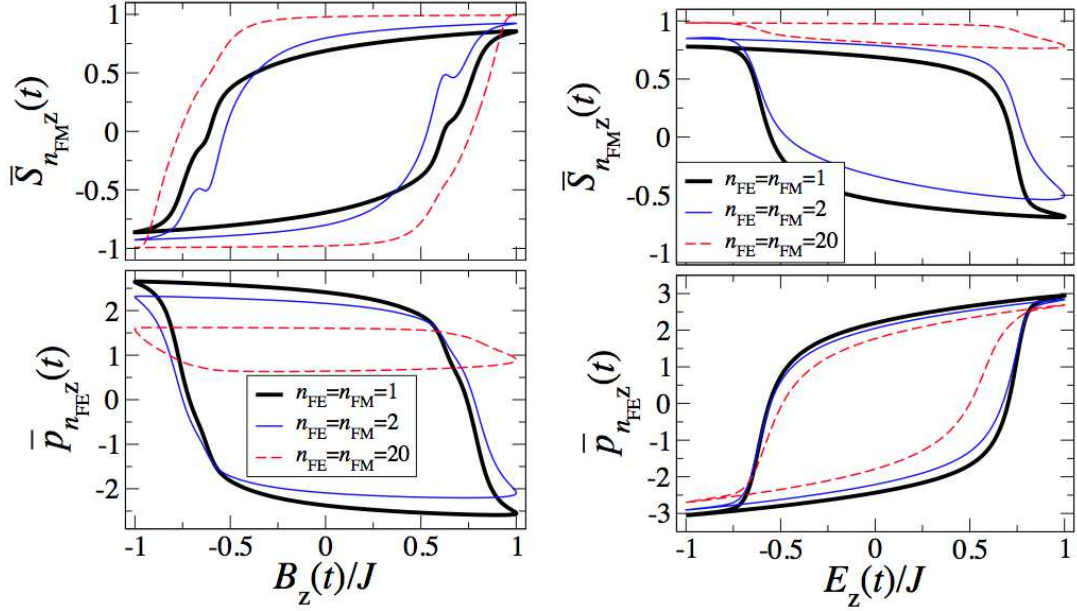}
\label{fig3}
\caption{The hysteresis loops of the renormalized polarization/magnetization of the first $n_i$ ($n_i$ =1, 2 or 20) FE and FM sites as a function of a harmonic  external magnetic field $B_z(t)$ (\emph{left}) or electric field $E_z(t) $  (\emph{right})  in the ME chain consisting of 200 polarization sites and 200 magnetic moments. The parameters are chosen to be the same as in Fig. \ref{fig2}.  }
\end{figure}

\section{Results and discussions}
The multiferroic response to external fields is investigated in  two cases: (I)  FE/FM chains with different lengths, $N_{FE}= N_{FM}$ = 2, 20, and 200; (II) the first $n_i$ ($n_i$ = 1, 2 and 20) polarization sites and magnetic moments around the interface of the long chain consisting of $N_{FE}= N_{FM} = 200$ \textcolor{black}{sites in total}. Temperature is set to zero. This ensures the FM-subsystem remains away from  its superparamagnetic limit and the FE-subsystem to be far from its phase transition temperature at \textcolor{black}{nearly} $T=10$~K for the whole time scale of simulations.

\textcolor{black}{Aiming at fast switching devices we apply harmonic fast oscillating magnetic and electric fields, as in Ref. \cite{SJHB}. We stress, however, that the switching by a magnetic field is not due to the resonance absorption of the magnetic field, as described in Refs. \cite{ACfields}. The strong magnetic field induces changes of the potential barrier compared with the height of it.}

In the case (I) with different sizes \textcolor{black}{of heterostructures}, the system is described via the reduced net magnetization $\vec{S}_{\Sigma}(t)$ and  reduced total polarization $\vec{p}_{\Sigma}(t)$
\begin{equation}
\displaystyle \vec{S}_{\Sigma}(t)=\frac{1}{N\ind{FM}}\sum_{i=1}^{N\ind{FM}} \vec{S}_i(t), ~~ \displaystyle \vec{p}_{\Sigma}(t)=\frac{1}{N\ind{FE}}\sum_{j=1}^{N\ind{FE}} \vec{p}_j(t).
\label{eq_9}
\end{equation}
Fig. \ref{fig2} shows the ME response of the total reduced polarization/magnetization to harmonic magnetic field  $B_z(t) = B_{0z} \cos \omega t$ and electric field $E_z(t) = E_{0z} \cos \omega t$, respectively. Clearly, the FE polarization (FM magnetization) is completely switched by the external electric (magnetic) field. Additionally, the hysteresis loops indirectly driven  by the applied field due to the ME coupling are presented. For the long ME chain, such as $N_{FE}= N_{FM} = 200$, the total FE polarization (magnetization) can not  be reversed by the magnetic (electric) field since the active area of the ME effect is  limited to the FE/FM interface only. A way around this problem is a multilayer stacking of alternating FE and FM structures coupled at their interfaces.

In order to investigate whether only the several first polarization (spin) sites in the long ME chain follow the magnetic (electric) field, the first $n_i$ ($n_i$= 1, 2 and 20) sites are extracted from the $N_{FE}=N_{FM} =200$ chain. For convenience, we define the following renormalized parameters

\begin{eqnarray}
\displaystyle &&\bar{\vec S}_{n_{FM}}(t)=\frac{1}{n_{FM}}\sum_{i=1}^{n_{FM}} \vec{S}_i(t), ~~
\displaystyle \bar{\vec{p}}_{n_{FE}}(t)=\frac{1}{n_{FE}}\sum_{j=1}^{n_{FE}} \vec{p}_j(t).
\label{eq_11}
\end{eqnarray}
The dynamical behavior is shown in Fig. 3.  Although the total FE polarization (FM magnetization)  driven indirectly by the external magnetic (electric) field, is not completely switched, it is indeed  observed
that the several first sites react strongly to the applied fields through the interface ME coupling.

\section{Conclusion}
 In summary, the dynamical multiferroicity in a one-dimensional FE/FM heterostructure is investigated by means of the coupled LLG and LKh method \textcolor{black}{as a function of the length of the heterostructure}. These results are supported by our kinetic Monte-Carlo simulations, not shown here
 for brevity. We find  that the FE polarization and the magnetization are \textcolor{black}{switchable for thin MF-chains (e.g. $N_{FE}=N_{FM}=1,2$ (Fig. \ref{fig2}) and up to $10$)} by applied  magnetic and electric fields, respectively. The \textcolor{black}{complete MF-}switching mediated by the multiferroic coupling is \textcolor{black}{not achievable for larger MF-chains (e.g. for $N_{FE}=N_{FM}=20$, as inferred from Fig. \ref{fig2})}.  Finally, we  mention that there are several additional effects that may affect the multiferroic dynamics. In Ref. \cite{SJHB}, the effect of the depolarizing fields and the effect of the induced electric (magnetic) fields are  discussed in some details.  These effects are quite small for the  BaTiO$_3$/Fe chain and for the chosen range of frequencies. The temperature- and the frequency-dependence of the dynamical response deserves further  considerations and is currently under study.

\end{document}